# Evaluating Journal Quality: A Review of Journal Citation Indicators and Ranking in Business and Management

**John Mingers (Corresponding author)**

Kent Business School, University of Kent, Canterbury CT7 2PE, UK

j.mingers@kent.ac.uk

01227 824008

**Liying Yang**

National Science Library, Chinese Academy of Sciences, 33 Beisihuan Xilu, Beijing 100190, China

yangly@mail.las.ac.cn

**Abstract:** Evaluating the quality of academic journal is becoming increasing important within the context of research performance evaluation. Traditionally, journals have been ranked by peer review lists such as that of the Association of Business Schools (UK) or though their journal impact factor (JIF). However, several new indicators have been developed, such as the h-index, SJR, SNIP and the Eigenfactor which take into account different factors and therefore have their own particular biases. In this paper we evaluate these metrics both theoretically and also through an empirical study of a large set of business and management journals. We show that even though the indicators appear highly correlated in fact they lead to large differences in journal rankings. We contextualise our results in terms of the UK's large scale research assessment exercise (the RAE/REF) and particularly the ABS journal ranking list. We conclude that no one indicator is superior but that the h-index (which includes the productivity of a journal) and SNIP (which aims to normalize for field effects) may be the most effective at the moment.

## 1. INTRODUCTION

The evaluation of research performance, whether at the level of individuals, departments, research groups or whole universities, is becoming ever more important and the results of exercises such as the UK's Research Excellence Framework (REF) have major consequences in terms of funding and individual academics' careers. The primary driver of an evaluation is an assessment of the quality of an individual research output, generally a journal paper. The evaluation can be done by peer review, as in the REF, or citations can be used as a proxy for quality – although they are really indicators of impact. The focus on quality of research has led to a focus on the quality of the publishing journal itself. There are several reasons for this: helping researchers decide where to target their papers;

competition between the journals; and in many cases illicitly using the quality of the journal as a proxy for the quality of the papers published in it.

Journal quality, in turn, can also be evaluated either by peer review or by citation indicators such as the journal impact factor (JIF). Peer review has been the primary form in the past for journal ranking lists such as that of the Association of Business Schools (ABS) (Association of Business Schools, 2010)[1]. Many of these lists for business and management are available from the Harzing website (2009). Some of these lists, such as ABS, are a hybrid in that they use citation indicators to inform the peer review.

The practice of judging a paper by the journal in which it is published has become endemic within large scale evaluations such as the UK's REF where huge numbers of papers need to be graded. Within business and management, in preparation for the 2014 REF, the ABS Guide was used by Schools to choose both papers and individual academics to be submitted, despite extensive criticism of the Guide from UK academics (Hussain, 2013; Mingers & Willmott, 2013; Walker et al., 2015; Willmott, 2011). It should be noted that the Business and Management REF Panel has repeatedly stated that they do not use and journal lists, and they have informally issued some data justifying this position, but this has not stopped the wholesale use of lists within business schools

This paper will discuss the results primarily within the UK context, but these large scale research evaluations also occur in Australia (Northcott & Linacre, 2010), New Zealand (Hicks, 2012) and Italy (Rebora & Turri, 2013).

These developments increase the importance of journal quality indicators, whether used in combination with peer review (as in the ABS list) or used instead of peer review. It is vital that the indicators available are accurate, robust, transparent and unbiased so that users, especially non-bibliometricians, can use then confidently (Wouters et al., 2015). For many years the journal impact factor (JIF) was the predominant journal metric despite considerable criticism, but recently there has been a spate of new ones including the Eigenfactor, the h-index, SJR and SNIP and it is important to understand how these differ from each other, and the degree of their validity(Moed, 2015; Straub & Anderson, 2010).

The purpose of this paper is to evaluate the indicators that are currently available in terms of these four criteria – accuracy, robustness, transparency and unbiasedness. It is evident that any metric has its own particular biases; that is, it will tend to favour certain kinds at the expense of others - that is after all the point of measuring something. Some of these biases will be explicit and indeed designed in. Others will be implicit, perhaps not recognized, and may be undesirable. The review will analyse

[1] In its latest incarnation it is called the "Academic Journal Guide 2015" and is published by the Chartered Association of Business Schools (2015)

the theory of the different indicators, looking for their explicit and implicit biases, and then test these observations on a sample of journals from the business and management area. There are four sections which cover: a review of the different indicators; methodology and data collection; empirical results and comparisons with the ABS list; and recommendations.

## 2. REVIEW OF JOURNAL CITATION INDICATORS

The use of citations to track the performance of journals was initiated by Garfield in 1955 (Garfield, 1955) and he established the first citation indexes (*Science Citation Index*)[2] and the company, the *Institute for Scientific Information (ISI)*. Although, prior to that, the first analysis of papers citing a journal's publications occurred in 1927 (Gross & Gross, 1927) and *Shepard's Citations* is a legal citing service started in 1873[3].

Mingers and Leydesdorff (2015b) and Cronin and Sugimoto (2014) provide reviews of the current field of scientometrics and bibliometrics as a whole. Waltman (2015) is a review of the literature on citation impact indicators including their use for journals although it does not include SNIP. There have also been several reviews of the use of indicators in research evaluation generally, some favourable (Abramo & D'Angelo, 2011; Moed, 2007; van Raan, 2005b) and some less so (Evidence Ltd, 2007; Leydesdorff & Milojević, 2015; Wouters et al., 2015).

### *2.1. Sources of citation data*

Before discussing the indicators it is important to examine the limitations of the sources of citation data. Traditionally, the major source of citation data was the Thomson Reuters ISI *Web of Science* (WoS) which is a specialised database covering all the papers in around 12,000 journals[4] . It also covers some conference proceedings[5] and is beginning to cover books[6]. In 2004 a similar database was started by Elsevier called *Scopus*[7]that covers 20,000 journals and also conferences and books. These databases capture the information directly from the journals that they cover and are generally reasonably accurate. A third source, since 2004, is *Google Scholar* (GS) based on the Google search engine. This works in an entirely different way by searching the Web to find relevant citations from whatever sources it can. Giles and Khabsa (2014) estimated that GS indexes about 100 million English-language documents.

Many studies have shown that the coverage of WoS and Scopus differs significantly between different fields, particularly between the natural sciences, where coverage is very good, the social sciences

[2] His initial purpose was not, in fact, to evaluate research but to help researchers search the literature.
[3] I would like to thank Terry Rose Wheeler for this information.
[4]http://wokinfo.com/essays/journal-selection-process/
[5]http://wokinfo.com/products_tools/multidisciplinary/webofscience/cpci/cpciessay/ (Cronin, 2001)
[6]http://thomsonreuters.com/book-citation-index/
[7]http://www.elsevier.com/online-tools/scopus/content-overview

where it is moderate and variable, and the arts and humanities where it is generally poor (HEFCE, 2008; Larivière et al., 2006; Mahdi et al., 2008; Moed & Visser, 2008). In contrast, the coverage of GS is generally higher, and does not differ so much between subject areas, but the reliability and quality of its data can be poor (Amara & Landry, 2012).There have also been many comparisons of WoS, Scopus and Google Scholar in different disciplines (Adriaanse & Rensleigh, 2013; Amara & Landry, 2012; Franceschet, 2010; García-Pérez, 2010; Harzing & van der Wal, 2008; Jacso, 2005; Meho & Rogers, 2008; Meho & Yang, 2007; Mingers & Lipitakis, 2010). The general conclusions of these studies are:

- That the coverage of research outputs, including books and reports, is much higher in GS, usually around 90%, and that this is reasonably constant across the subjects. This means that GS has a comparatively greater advantage in the non-science subjects where Scopus and WoS are weak.
- Partly, but not wholly, because of the coverage, GS generates a significantly greater number of citations for any particular work. This can range from two times to five times as many. This is because the citations come from a wide range of sources, not being limited to the journals that are included in the other databases.
- However, the data quality in GS is very poor with many entries being duplicated because of small differences in spellings or dates and many of the citations coming from a variety of non-research sources. With regard to the last point, it could be argued that the type of citation does not necessarily matter – it is still impact.
- There is also an issue about normalising Google Scholar data – see the later section on normalisation (Bornmann et al., 2016).
- Even with the specialised databases (WoS and Scopus) there are still issues around error-correction and disambiguation of common names.

## *2.2. Basic Journal Citation indicators*

After a paper has been published it may be cited in later papers. The basic unit is the number of times a paper, or a collection of papers, has been cited over a particular time window. In the case of journals, it will be the number of citations to documents of a particular type that were published in the journal over a specific period. Thus, it is generally necessary to specify two time periods, which may be the same, one for the cited papers and one for the citing papers.

### *Total Citations*

The most basic citation metric is simply the total number of citations received by papers in a journal in the relevant time periods. This measure is accurate, robust and transparent. However, it is not unbiased since it is dependent on many factors, some of which do not relate to quality and should therefore be controlled for. The first is the number of papers that the journal publishes. Clearly the more papers published per year the more citations that will accrue but it could be argued that this degree of productivity is not the same as quality or impact. This leads to the mean or average number of citations. The second is that different research fields have very different citation practices and generally the sciences have a much greater citation density than the social sciences or humanities, for example molecular biology was found to have citation rates ten times greater than computer science (Iglesias & Pecharromán, 2007). But there may be significant differences even within a multidisciplinary field such as business and management (Mingers & Leydesdorff, 2015a). This is also related to the average number of authors for a paper – in lab-based science particularly, papers[8] can have many authors effectively increasing the overall number of citations. This is a major problem which makes it very difficult to compare journals across different research fields without some form of field or source normalization.

A third factor is the quality of the citation itself – should all citations, from whatever source, be considered equal or is a citation from a highly prestigious journal worth more than one from an obscure journal?

*Average number of citations per paper (CPP or IPP)*

Dividing the total citations by the number of papers generating them gives the citations per paper (CPP) (van Raan, 2003) or impact per paper (IPP) (Moed, 2010b). This completely normalizes for the number of papers but does not normalize for field. Another limitation is that citation data is always very highly skewed (Seglen, 1992) and so mean rates may well be distorted by extreme values. This has led to the use of non-parametric measures such as the number or proportion of highly-cited papers (Bornmann et al., 2013; Leydesdorff & Bornmann, 2011; Leydesdorff et al., 2011).

*Journal impact factor (JIF)*

The JIF was the first, and is the most well-known journal metric. This was originally developed by Garfield and Sher (1963) as a way of choosing journals to include in their newly-created science citation index (SCI). It is simply a two year mean citations per paper based on the number of citations in year *t* to papers published in the previous two years. JIF is published every year for all the journals that are included in Thompson Reuters' *Web of Science* and is viewed as highly influential. Garfield recommended that it should be used in combination with another metric, the "cited half-life" which measures how long citations last for – it is the median age of papers cited by a journal in a particular JCR year. WoS also calculates the immediacy index which is a 1-year JIF, i.e., the mean citations in year t to papers published in year t. The JIF has several limitations (Glänzel & Moed, 2002; Harzing & Van der Wal, 2009):

---

[8] As an extreme, although not uncommon, example the papers announcing the discovery of the Higgs boson (Aad et al., 2012) had 3,000 authors and currently has 6,700 citations.

- JIF depends heavily on the research field. For instance, in 2013 the top journal in cell biology had a JIF of 36.5 and *Nature* has one of 42.4 while the top journal in management, *Academy of Management Review,* has a JIF of only 7.8 and many are less than 1.
- The two-year window. This is a very short time period for many disciplines, especially given the lead time between submitting a paper and having it published which may itself be two years. The 5-year JIF is better in this respect (Campanario, 2011).
- There is a lack of transparency in the way the JIF is calculated and this casts doubt on the results. Brumback (2008) studied reviews journals and could not reproduce the appropriate figures. It is highly dependent on which types of papers are included in the denominator. Pislyakov (2009) compared JIFs calculated from WoS and Scopus data and found significant differences mainly due to the different coverage of the databases. The situation may be improved more recently.
- It is possible for journals to deliberately distort the results by, for example, publishing many review articles which are more highly cited; publishing short reports or book reviews that get cited but are not included in the count of papers; publishing yearly overviews of the research published in the journal or pressuring authors to gratuitously reference excessive papers from the journal (Lowry et al., 2013; Moed, 2000; Wilhite & Fong, 2012).

*The h-index*

This is a relatively new indicator proposed by Hirsch (2005) that has generated a huge amount of interest. It can be used for journals, individual researchers, or departments. We will only summarise the main advantages and disadvantages, for more detailed reviews see (Alonso et al., 2009; Bornman & Daniel, 2005; Costas & Bordons, 2007; Glänzel, 2006; Norris & Oppenheim, 2010) and for mathematical properties see Glänzel(2006) and Franceschini and Maisano (2010).

The h index is defined as: "a scientist has index $h$ if $h$ of his or her $N_p$ papers have at least $h$ citations each and the other $(N_p – h)$ papers have <= $h$ citations each" Hirsch (2005, p. 16569).

Thus $h$ is the top $h$ papers of a collection that all have at least $h$ citations. The novel property of $h$ is that in one number it summarizes both impact, in terms of citations, and productivity in terms of number of papers. It thus lies somewhere between CPP, which ignores productivity, and total cites which is heavily dependent on productivity. The $h$ papers are generally called the h-core. The h-index ignores all papers outside the h-core, and also ignores the actual number of citations received by the h-core papers. The strengths of the h-index are:

- It combines both productivity and impact in a single measure that is easily understood and very intuitive.

- It is easily calculated just knowing the number of citations either from WoS, Scopus or Google Scholar (GS). Indeed, all three now routinely calculate it.
- It can be applied at different levels – researcher, journal or department.
- It is objective and a good comparator within a discipline where citation rates are similar.
- It is robust to poor data since it ignores the lower down papers where the problems usually occur. This is particularly important if using GS.

However, many limitations have been identified including some that affect all citation based measures (e.g., the problem of different scientific areas, and ensuring correctness of data), and a range of modifications have been suggested (Bornmann et al., 2008).

- The metric is insensitive to the actual number of citations so two journals could have the same h-index but very different total citations. The g-index (Egghe, 2006) has been suggested as a way of compensating for this.
- The h-index is strictly increasing and strongly related to the time the publications have existed. This biases it against newer journals. It is possible to time-limit the h-index, for example *Google* Metrics uses a 5-year h-index (Jin et al., 2007).
- The h-index is field dependent and so should be normalized in some way. Iglesias and Pecharroman (2007) constructed a table or normalisation factors for 21 different scientific fields.
- The h-index is dependent on or limited by the total number of publications and this is a disadvantage for journals which are highly cited but for a relatively small number of publications (Costas & Bordons, 2007). It will thus tend to favour journals that publish many papers against those with a small number of high quality papers. This can be seen clearly in the empirical results.

There have been many comparisons of the h-index with other indicators (Bornmann & Daniel, 2007; Lehmann et al., 2006; van Raan, 2005a). Generally, such comparisons show that the h-index is highly correlated with other bibliometric indicators, but more so with measures of productivity such as number of papers and total number of citations, rather than with citations per paper which is more a measure of pure impact (Alonso et al., 2009; Costas & Bordons, 2007; Todeschini, 2011). There have been several studies of the use of the h-index in business and management fields such as information systems (Oppenheim, 2007; Truex III et al., 2009), management science (Mingers, 2008; Mingers et al., 2012), consumer research (Saad, 2006), marketing (Moussa & Touzani, 2010) and business (Harzing & Van der Wal, 2009).

### *2.3. Normalisation*

One of the main principles of bibliometric analysis is that citations indicators from different academic fields should not be compared directly with one another because of the major differences in citation density across fields. It is also desirable to consider differences in publication type, for example, journals with large number of review papers, which are highly cited, or editorials or book reviews which generate citations but which might not be counted as papers. In this paper we will discuss three approaches to normalisation – field normalisation, percentile normalisation and citing-side or source normalisation – for empirical analysis see Waltman and Marx (2015) and Waltman and van Eck (2013).

### *Field normalisation*

Field normalisation means comparing the number of citations for a paper or journal, whether in absolute or average form, with the expected number of citations within the appropriate research field. For example, van Leeuwen and Moed (2002) developed a citation impact indicator that normalizes for field, publication year, and document type[9]. This works by comparing the number of citations received by a paper with the mean number of citations of similar papers across all journals in the field. The main problem is determining an appropriate field, and corresponding journals, for each paper. This is generally implemented within WoS and the WoS field lists are used. This approach forms the basis of the well-established methodology for evaluating research centres developed by the *Centre for Science and Technology Studies* (CWTS) at Leiden University known as the crown indicator or Leiden Ranking Methodology (LRM) (Moed, 2010c; van Raan, 2005c). The problems with this approach are that the WoS field categories are ad hoc, with no systemic basis (Leydesdorff & Bornmann, 2014; Mingers & Leydesdorff, 2015a) and that it is difficult to cope with inter-disciplinary papers or journals (Rafols et al., 2012) [10].

This form of normalisation is particularly difficult with Google Scholar data as there are no field lists provided in GS. One attempt at normalising GS data has been made (Bornmann et al., 2016) but it is very time-consuming and messy, and the results are not that reliable.

### *Citing-side or source normalisation*

An alternative method, originally suggested by Zitt and Small (2008) in their "audience factor", is to consider the source of citations – that is the reference lists of citing papers. The assumption is that high density fields will have large reference lists and low density fields short ones. This approach is also known as "citing-side approach" (Zitt, 2011), fractional counting of citations (Leydesdorff & Opthof, 2010) and a priori normalisation (Glänzel et al., 2011). It is the basis of the SNIP metric to be discussed later (Moed, 2010b).

This approach is different in that the reference set of journals is not defined in advance according to a WoS category, but is rather determined at the time as the set of all papers or journals that have cited the journal in question. Each evaluated journal will therefore have its own specific set of citing journals thus avoiding problems with outdated and ad hoc WoS categories. The disadvantage is that the journals are not being compared against the same benchmark set.

---

[9] The origin for this is Moed et al (1995)

[10] In their latest 2015 university ranking, CWTS do not use WoS field categories but instead have developed a set of fields algorithmically (see http://www.leidenranking.com/methodology/fields).

### *2.4. Second generation indicators*

In recent years several new, and more complex, indicators have been developed to take into account concerns about normalisation and the relative prestige of citing journals. Some of these indicators are specific to particular data sources, e.g., the Eigenfactor in WoS, and SNIP and SJR in Scopus

*Indicators measuring the prestige of citations: Eigenfactor and SJR*

The idea of these indicators is that having a paper cited in a very high-quality or prestigious journal such as *Nature* or *Science* is worth more than a citation in an obscure journal. The indicators all work on a recursive algorithm similar to that of Google's PageRank for web pages. The first such was developed by Pinsky and Narin (1976) but that had calculation problems. Since then, Page et al. (1999) and Ma (2008) have an algorithm based directly on PageRank but adapted to citations; Bergstrom (2007) has developed the Eigenfactor which is implemented in WoS; and Gonzalez-Pereira et al (2010) have developed SCImago Journal Rank (SJR) which is implemented in Scopus. We will use the latter two.

The Eigenfactor essentially measures the relative frequency of occurrence of each journal in the network of citations, and uses this as a measure of prestige. It explicitly excludes journal self-citations unlike most other indicators. Its values tend to be very small, for example the largest in the management field is *Management Science* with a value of 0.03 while the $20^{th}$ is 0.008, figures which are not easily interpreted. The Eigenfactor is based on the total number of citations and so is affected by the total number of papers published by a journal. A related metric, also in WoS, is the Article Influence Score (AIS) which is the Eigenfactor divided by the proportion of papers in the database belonging to a particular journal over five years. This is therefore similar to a 5-year JIF but normalized so that a value of 1.0 shows that the journal has average influence; values greater than 1.0 show greater influence.

The SJR works iteratively in a similar way to the Eigenfactor but its value is normalized by the total number of citations in the citing journal for the year in question. It works in two stages: firstly calculating an un-normalized value iteratively based on three components – a fixed amount for being included in Scopus, a value dependent on the number of citations received, and the prestige of the sources. There are a number of seemingly-arbitrary weightings in the formula. This value is then normalized by the number of published articles and adjusted to give an "easy-to-use" value. The currently implemented version of SJR in Scopus has a further refinement (Guerrero-Bote & Moya-Anegón, 2012) in that the "relatedness" of the citing journal is also taken into account. A significant problem with this metric (and with SNIP) is that its results are not reproducible outside of its actual production, for example by other researchers.

There are several limitations of these 2nd generation measures: the values for "prestige" are difficult to interpret as they are not a mean citation value but only make sense in comparison with others; they are still not normalized for field (Lancho-Barrantes et al., 2010); and the fields themselves are open to disagreement (Mingers & Leydesdorff, 2015a).

*Source-normalized indicator: SNIP*

Another 2nd generation metric is SNIP (Moed, 2010a) – source normalized impact per paper - which is only available within Scopus. This normalizes for different fields using the citing-side form of normalisation. It firstly calculates a 3-year IPP (effectively a 3-year JIF). It then calculates the "database citation potential" DCP for the particular journal by finding all the papers in year *n* that cite papers from the journal in the preceding ten years and calculating the mean of the number of *references* in those papers to papers within the database – i.e., Scopus. Next, the DCP for *all* journals in the database is calculated and the median of these values noted. The DCP for the journal is then divided by the median to relativize it to journals as a whole creating a relative DCP (RDCP). If this value is above 1 then the field has greater citation potential; if it is less than 1 the field has lower citation potential. Finally, SNIP = IPP/RDCP. If the field is high density then RDCP will be above 1 and the IPP will be reduced and vice versa if the field is low density. The currently implemented version of SNIP has two changes (Waltman et al., 2013): the DCP is calculated using the harmonic mean rather than the arithmetic mean, and the relativisation of the DCP is now dropped.

This is an innovative measure both because it normalizes for both number of publications and field, and because the set of reference journals are specific to each journal rather than being defined beforehand somewhat arbitrarily. Moed (2010a) presents empirical evidence from the sciences that the subject normalisation does work even at the level of pairs of journals in the same field. But, it is complex and rather opaque and criticisms have been levelled by Leydesdorff and Opthof (2010) and Mingers (2014).

*Percentile-based indicator – I3*

This approach aims to overcome the statistical problems of using means with highly skewed data. This uses WoS field categories to establish percentile ranks (PR) in terms of the number of citations necessary for a paper to be in the top 1%, 5%, 10% … of papers published in the field (Leydesdorff, 2012). The target set of papers for a journal are then all evaluated to see which PR they fall into and the proportions falling into each one are calculated. These can then be compared so that, for example, a journal with 5% of its papers having more citations than the top 1% in its field is above average.

Based on this form of normalisation, a metric has been developed as an alternative to the journal impact factor (JIF) called I3 (Leydesdorff & Bornmann, 2011). Instead of multiplying the percentile ranks by the proportion of papers in each class, they are multiplied by the actual numbers of papers in each class thus giving a measure that combines productivity with citation impact. This indicator is not available in any of the databases and so we will not be able to include it in the empirical investigation.

Table 1 summarises the main characteristics and the advantages and disadvantages of the indicators. It also shows typical values for a high density field (cell biology) and the management field.

## 3. Methodology and Data

We wished to compare the various indicators empirically on a sample of business and management journals and then compare the results with the ABS journal ranking. One of the problems is that the indicators are not all available from the same source – the JIF and Eigenfactor come from the WoS; the h-index from the Scimago website; and the SJR and SNIP from Scopus. Clearly this introduces problem of consistency as the databases do not cover the same set of journals and therefore have differences in the numbers of citations (Leydesdorff et al., 2014). However, from a practical viewpoint in terms of using these indicators either by themselves or as part of creating a journal list such as ABS, we have to accept what is available and so it is the data as it stands that we have analysed.

Data was collected from these three sources for the years 2012 and 2013. The analysis we present is based on 2013 but checks with 2012 showed no major inconsistencies. The data was validated, especially in terms of ensuring consistency of journal title and ISSN, and a small number of outliers were removed. The full dataset contains the following variables – note that there are many less journals classified as business and management in WoS. This meant that the statistical analyses were often restricted by the smaller WoS size.

| Metric | Description | Advantages | Disadvantages | Maximum values for: a) cell biology b) management | Normalizes for: | | |
|---|---|---|---|---|---|---|---|
| | | | | | No of papers | Field | Prestige |
| Impact factor (JIF) (WoS) | Mean citations per paper over a 2 or 5 year window. Normalized to number of papers. Counts citations equally | Well-known, easy to calculate and understand. | Not normalized to discipline; short time span; concerns about data and manipulation | From WoS<br>a) 36.5<br>b) 7.8 | Y | N | N |
| Eigenfactor and article influence score (AIS) (WoS) | Based on PageRank, measures citations in terms of the prestige of citing journal. Not normalized to discipline or number of papers. Correlated with total citations. Ignores self-citations. AI is normalized to number of papers, so is like a JIF<br>5-yr window | The AI is normalized to number of papers. A value of 1.0 shows average influence across all journals | Very small values, difficult to interpret, Eigenfactor not normalized | From WoS<br>Eigenfactor:<br>a)0.599<br>b)0.03<br><br>AI:<br>a) 22.2<br>b) 6.56 | N<br><br>Y | N<br><br>N | Y<br><br>Y |
| SJR and SJR2 (Scopus) | Based on citation prestige but also includes a size normalisation factor.<br>SJR2 also allows for the closeness of the citing journal. 3-year window | Complex calculations and not easy to interpret. Not field normalized | Normalized number of papers but not to field so comparable to JIF. Most sophisticated indicator | From Scimago<br>a) 22.43<br>b)15.3 | Y | N | Y |
| h-index (Scimago website and Google Indicators) | The h papers of a journal that have at least h citations. Can have any window – Google indicators uses 5-year | Easy to calculate and understand. Robust to poor data | Not normalized to number of papers or field.<br>Not pure impact but includes volums | From Google Indicators<br>h5:<br>a) 223<br>b) 72 | N | N | N |
| SNIP<br><br>Revised SNIP (Scopus) | Citations per paper normalized to the relative database citation potential, that is the mean number of references in the papers that cite the journal | Normalizes both to number of papers and field. | Does not consider citation prestige.Complex and difficult to check.Revised version is sensitive to variability of number of references | From CWTS<br>a) 7.9<br>b) 6.19 | Y | Y | N |
| I3 | Combines the distribution of citation percentiles with respect to a reference set with the number of papers in each percentile class | Normalizes across fields. Does not use the mean but is based on percentiles which is better for skewed data | Needs reference sets based on pre-defined categories such as WoS | Not known | N | Y | N |

**Table 1 Characteristics of Indicators for Measuring Journal Impact (Based on (Mingers & Leydesdorff, 2015b) Table 2)**

Table 2 shows the descriptive statistics with the following points of note:

- The different values of N reflect the different sources, i.e, WoS or Scopus or Scimago.
- There are widely different values from a mean of 0.00317 for the Eigenfactor to 129.8 for 3-year total citations.
- The variables, especially those involving un-normalized values of citations, are all highly skewed (the critical point is generally taken to be $[6/N]^{1/2}$).
- All of the variables except the immediacy index are some form of citation metric. We have not included the citation half-life as it is time-based.

| Variable | N | Mean | StDev | Median | Maximum | Skewness |
|---|---|---|---|---|---|---|
| H index | 1279 | 16.884 | 23.381 | 7.000 | 182.000 | 2.93 |
| 3-year Total Cites | 1279 | 129.80 | 330.18 | 33.00 | 6028.00 | 8.04 |
| IF | 426 | 1.2751 | 1.0790 | 0.9875 | 7.8170 | 2.22 |
| Total cites | 426 | 1619 | 3047 | 696 | 26370 | 4.48 |
| 5-Year IF | 426 | 1.6532 | 1.6004 | 1.2780 | 10.1540 | 1.92 |
| Immediacy Index | 426 | 0.2433 | 0.2674 | 0.1570 | 2.7140 | 3.09 |
| Eigenfactor Score | 426 | 0.003173 | 0.006325 | 0.001320 | 0.063680 | 5.90 |
| Article Influence Score | 426 | 0.7644 | 1.0954 | 0.4535 | 9.0960 | 3.91 |
| SNIP | 1022 | 0.8588 | 0.7743 | 0.6545 | 5.9210 | 2.08 |
| IPP | 1022 | 0.9828 | 1.1323 | 0.6215 | 8.7220 | 2.67 |
| SJR | 1134 | 0.7026 | 1.3002 | 0.2975 | 18.4400 | 6.03 |

**Table 2 Descriptive Statistics for the Journal Indicators**

The next step was to look at the correlations between the various indicators.

**Figure 1 Correlation Plots between Indicators**

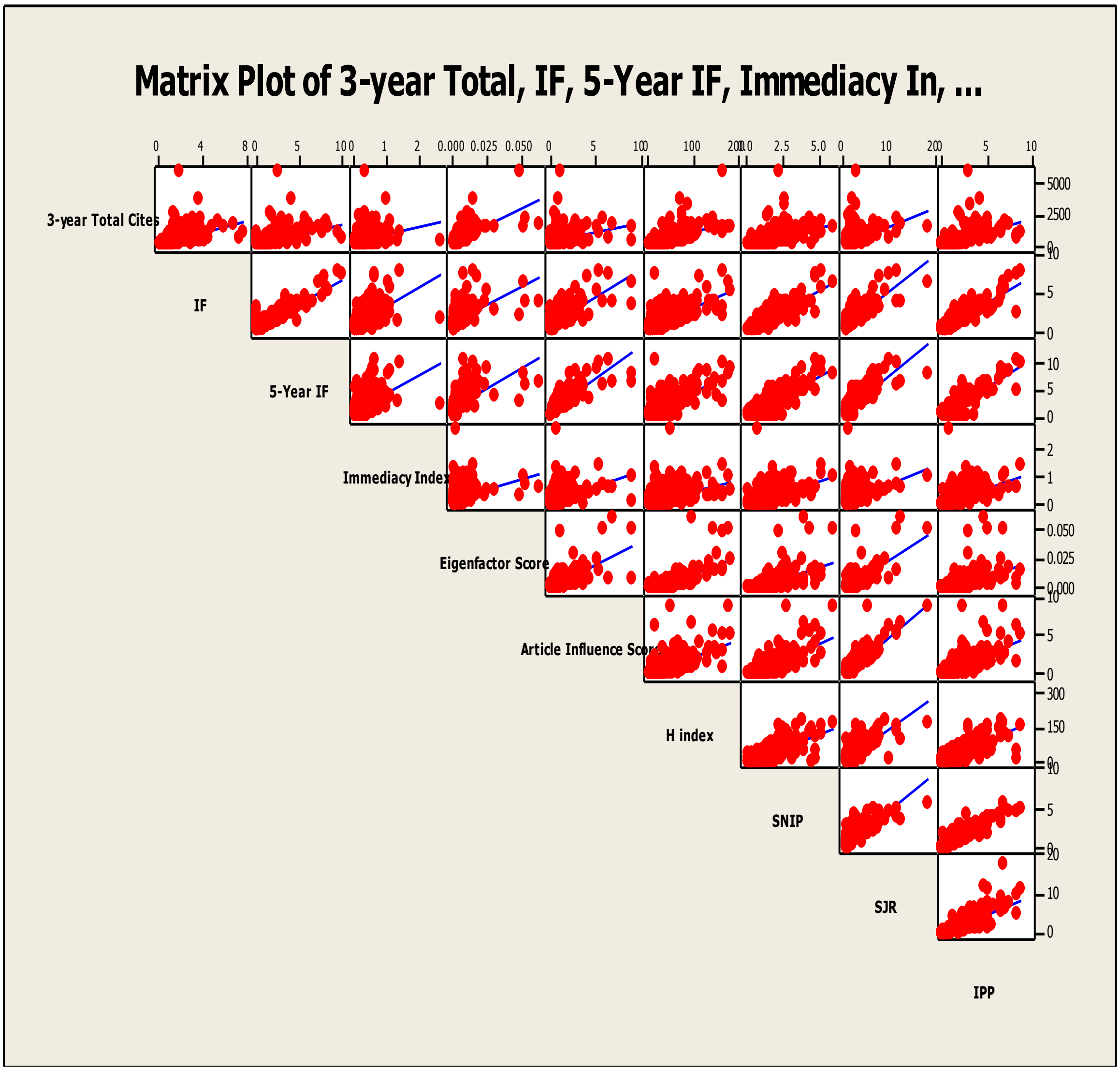

We can see that several combinations, especially the various versions of the impact factor, have strong linear relationships as we would expect. We can therefore calculate the Pearson's correlation coefficient as in Table 3.

| | Tot.Cit | IF | 5-IF | Immed. | Eigen. | AIS | H-Ind. | SNIP | SJR |
|---|---|---|---|---|---|---|---|---|---|
| IF | 0.535 | | | | | | | | |
| 5-Year IF | 0.529 | 0.920 | | | | | | | |
| Immediacy Ind. | 0.353 | 0.600 | 0.551 | | | | | | |
| Eigenfactor | 0.734 | 0.559 | 0.596 | *0.352* | | | | | |
| Article Infl. | 0.371 | 0.740 | 0.824 | *0.413* | 0.719 | | | | |
| H index | 0.743 | 0.715 | 0.764 | *0.468* | 0.737 | 0.657 | | | |
| SNIP | 0.601 | 0.853 | 0.857 | 0.523 | 0.615 | 0.745 | 0.800 | | |
| SJR | 0.553 | 0.806 | 0.835 | *0.462* | 0.780 | 0.906 | 0.779 | 0.807 | |
| IPP | 0.639 | 0.917 | 0.908 | 0.552 | 0.559 | 0.649 | 0.814 | 0.924 | 0.827 |

Table 3. Correlation Coefficients (above 0.9 are greyed, below 0.5 are italicized)

As expected, all of the coefficients are statistically significant at the 0.01 level. Those based on citations per paper (IF, 5-year IF, IPP) have very high correlations as do article influence and SJR, and SNIP and IPP. The latter one is interesting as SNIP is a normalized version of IPP but this shows that either the normalization does not work well, or that that journals show similar referencing patterns. Also of note is that the immediacy index has the lowest correlations with the other indicators. We should be careful, however, not to conclude from high correlations that the indicators are all measuring the same thing – West et al (2010) have shown that even in a sample where the Eigenfactor and total citations were correlated at the 0.995 level there were still significant differences in journal rankings between them.

## 4. Analysis of the Results

### *4.1. Principal Components*

Given the interesting pattern of correlations, it is useful to conduct a principal components analysis to look at the relationships between the variables. Figure 2 is a plot of the first two component loadings. PC1 Does not discriminate well between the indicators although those normalized for number of papers have higher values. PC2 distinguishes clearly between these types of indicators with those un-normalized having positive values. We have not included the immediacy index as this is something of an outlier as it is a very short-term 1-year JIF and so may not be appropriate for business and management where citation rates are slow in comparison with science.

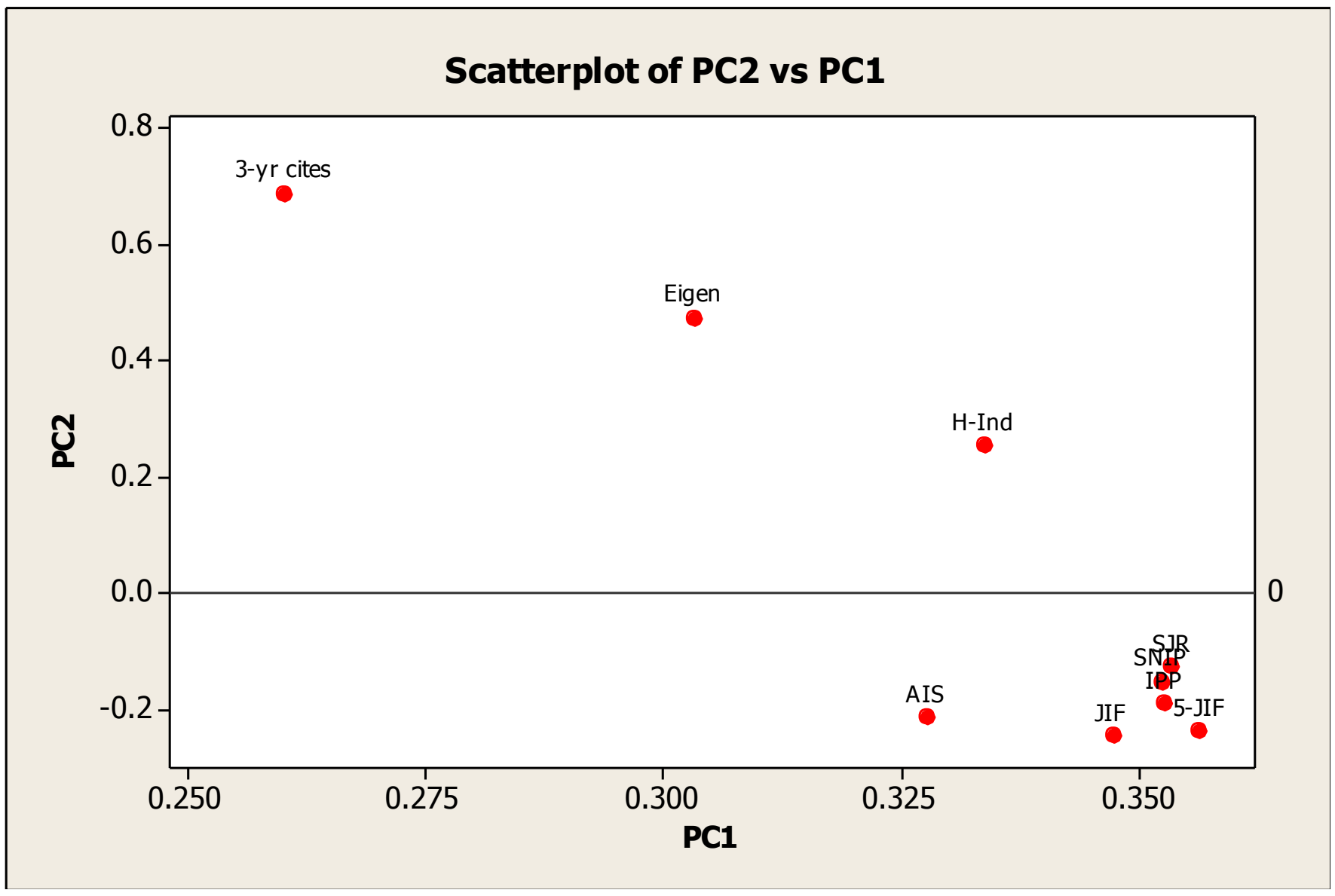

**Figure 2 Plot of Principal Component Loadings 1 and 2**

From this plot we can see that the indicators fall into several groups plus some outliers. The main group are all those which normalize citations for the number of papers, that is JIF, 5-JIF, IPP, SNIP and SJR. At this level of aggregation the fact that SJR allows for prestige and SNIP for field effect does not seem evident. Close to this group is the article influence score (AIS) which is the Eigenfactor normalized for the number of papers. Towards the top are the 3-year total cites and also the Eigenfactor, neither of which normalize for papers. The h-index comes between these two groups which seems appropriate.

There are two further points of note. First, SJR is closer to the other non-prestige indicators than it is to AIS which is the other prestige based metric, and it is far away from the Eigenfactor suggesting perhaps that it is not measuring prestige in the same way. Second, SNIP is very close to the impact factors suggesting that the source normalization is not having much effect. This could be because the citation practices within business and management do not differ greatly but there is evidence against this – Mingers and Leydesdorff (2015a), in an analysis of journal cross-citation rates, identified six different sub-fields where the citation rates differed significantly, from a CPP of 32.5 in marketing, IB, strategy and IS, to 11.8 in operational research and operations management.

Figure 3 is a plot of PC2 vs PC3. Here, we can see a clear differentiation between those that take prestige into account (SJR, Eigenfactor and AIS) and those that do not. This is perhaps the clearest categorisation of the indicators: NE, total cites and no prestige; SE, cites per paper and no prestige; SW, cites per paper and prestige; and NW, total cites and prestige.

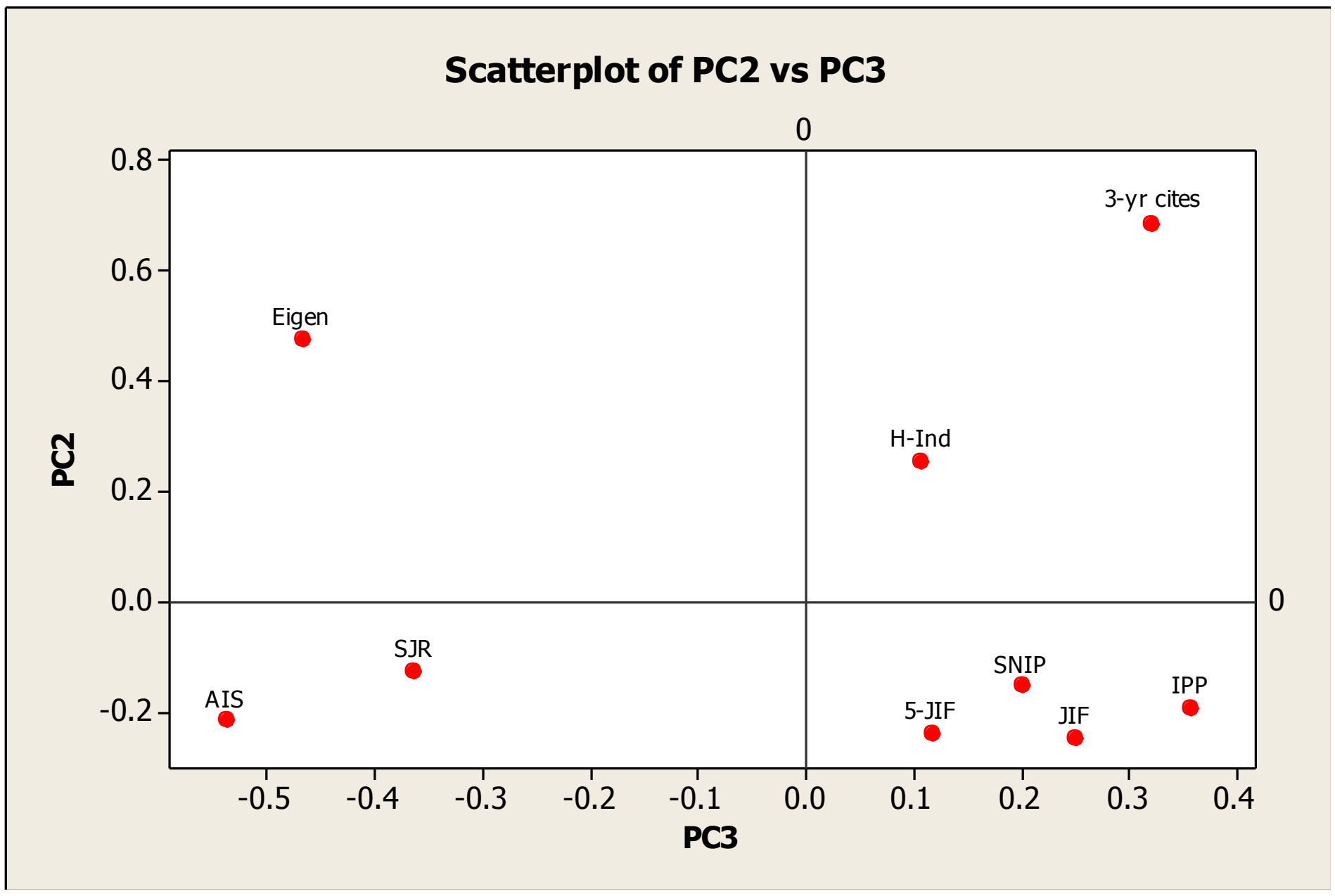

**Figure 3Plot of Principal Component Loadings 2 and 3**

Overall, the empirical results suggest that the theoretical differences between indicators can be detected at an aggregate level in the empirical results. We now turn to the practical results in terms of the rankings of journals using these indicators.

### *4.2. Rank ordering of journals*

The questions to be considered in this section are the extent to which the different indicators rank order journals differently, and whether these differences reflect the theoretical differences described above. Presenting the results is difficult as we are comparing so many indicators at the same time - most studies compare only one or two. For example, Leydesdorff, and Bornmann (2011) compared the integrated impact indicator (a percentile based metric) with the journal impact factor concluding that the percentile approach had advantages. Falagas et al (2008) compared the JIF with the SJR concluding that both had advantages and disadvantages. Bollen et al (2006) compared JIF with a PageRank algorithm concluding that they were measuring rather different things – popularity as opposed to prestige – and suggested the Y-factor as a combination of the two. Fersht (2009) discussed the JIF and the Eigenfactor and concluded that the Eigenfactor was strongly related to total citations, and that individual scientists should be judged by the h-index rather than journal indicators. In two other comparison studies, Glanzel et al (2011) looked at normalizing journal indicators and compared a priori normalization by using fractional counting of citations with a posteriori normalization of the JIF and suggested that both were useful approaches. .

In Table 4 we have ranked the journals using the different indicators and we have also summed the ranks to give an overall ranking. The Table is ordered in terms of this summed rank. We have also shown the 2009 ABS journal list category (where available)[11] and the field according to ABS as well as total citation over a 3 year period and total documents published per year. In using the sum of the ranks to select and order the journals, we are thereby biasing this table towards those journals that do well across the board, that is that do not do badly on any one metric. There may well be journals that do very well on the majority but do particularly badly on one or two indicators and they would not appear. We will see examples of this below.

First, we can notice the following general points:

[11] We use the 2009 version as that was current at the time of the data. The latest version, 2014, has not, unfortunately, been made available in a downloadable form and so could not be used. Hopefully ABS will make it available for research purposes in the future.

- The journals come from a range of different ABS sub-fields and they also have a wide range of total papers published - from 12 to 709, and citations - from 140 to 2029 so there is no obvious overall pattern.
- The vast majority are classified by ABS as 4* journals but five are only classified as 3* and three are not even included in ABS – *International Organization, Academy of Management Perspectives* and the *Journal of Cleaner Production.* We will show more detailed comparisons with ABS below.

| Standard Title | Field | ABS 2009 1* to 4* | Total Docs. | 3-year Total Cites | H index rank | IF Rank | 5-Year IF | IPP Rank | SNIP Rank | SJR Rank | Eigen Rank | AI Score Rank | Sum of Ranks | Sum ranked |
|---|---|---|---|---|---|---|---|---|---|---|---|---|---|---|
| Journal of Finance | FINANCE | 4 | 69 | 1420 | 2 | 4 | 7 | 6 | 1 | 1 | 3 | 2 | 26 | 1 |
| Academy of Management Review | GEN MAN | 4 | 43 | 987 | 5 | 1 | 2 | 1 | 2 | 4 | 16 | 6 | 37 | 2 |
| Academy of Management Journal | GEN MAN | 4 | 80 | 1384 | 1 | 6 | 3 | 8 | 13 | 6 | 6 | 7 | 50 | 3 |
| Journal of Management * | GEN MAN | 4 | 71 | 1533 | 12 | 3 | 5 | 4 | 5 | 8 | 9 | 9 | 55 | 4 |
| Journal of Financial Economics | FINANCE | 4 | 158 | 2029 | 8 | 13 | 13 | 14 | 9 | 3 | 2 | 5 | 67 | 5 |
| Journal of Marketing | MKT | 4 | 48 | 1094 | 7 | 10 | 8 | 9 | 8 | 11 | 15 | 14 | 82 | 6 |
| Review of Financial Studies | FINANCE | 4 | 70 | 1626 | 21 | 17 | 9 | 20 | 12 | 2 | 1 | 3 | 85 | 7 |
| MIS Quarterly | INFO MAN | 4 | 68 | 1771 | 10 | 5 | 4 | 5 | 3 | 13 | 26 | 21 | 87 | 8 |
| Strategic Management Journal | STRAT | 4 | 186 | 1264 | 3 | 27 | 11 | 17 | 18 | 7 | 8 | 16 | 107 | 9 |
| Organization Science | ORG STUD | 4 | 95 | 1407 | 9 | 11 | 16 | 23 | 34 | 9 | 7 | 11 | 120 | 10 |
| Journal of Operations Management | OPS & TECH | 4 | 37 | 916 | 15 | 8 | 6 | 7 | 15 | 16 | 50 | 33 | 150 | 11 |
| Personnel Psychology * | HRM&EMP | 4 | 44 | 462 | 36 | 7 | 12 | 15 | 19 | 17 | 42 | 13 | 161 | 12 |
| Journal of Accounting and Economics | ACCOUNT | 4 | 31 | 682 | 28 | 34 | 23 | 10 | 11 | 10 | 33 | 19 | 168 | 13 |
| Journal of International Business Studies | IB&AREA | 4 | 49 | 1026 | 14 | 15 | 15 | 18 | 28 | 24 | 21 | 35 | 170 | 14 |
| Journal of Consumer Research | MKT | 4 | 81 | 1091 | 18 | 37 | 21 | 22 | 22 | 15 | 12 | 24 | 171 | 15 |
| Journal of Management Studies | GEN MAN | 4 | 66 | 1082 | 26 | 20 | 17 | 19 | 23 | 31 | 20 | 29 | 185 | 16 |
| Journal of Business Venturing | ENT-SMBUS | 4 | 64 | 753 | 23 | 22 | 24 | 12 | 14 | 27 | 52 | 34 | 208 | 17 |
| Journal of Marketing Research | MKT | 4 | 54 | 1030 | 19 | 42 | 36 | 35 | 43 | 12 | 10 | 18 | 215 | 18 |
| Journal of Organizational Behavior* | ORG STUD | 4 | 83 | 875 | 22 | 23 | 22 | 24 | 33 | 39 | 29 | 32 | 224 | 19 |
| Research Policy* | SOC SCI | 4 | 149 | 1721 | 11 | 45 | 31 | 30 | 27 | 48 | 13 | 49 | 254 | 20 |
| Organizational Research Methods* | ORG STUD | 4 | 28 | 449 | 77 | 18 | 14 | 25 | 50 | 20 | 36 | 17 | 257 | 21 |
| Management Science | OR&MANSCI | 4 | 166 | 1427 | 6 | 49 | 45 | 58 | 44 | 32 | 5 | 23 | 262 | 22 |
| Journal of the Academy of Marketing Science* | MKT | 4 | 52 | 772 | 17 | 19 | 25 | 21 | 32 | 40 | 61 | 52 | 267 | 23 |

| | | | | | | | | | | | | | |
|---|---|---|---|---|---|---|---|---|---|---|---|---|---|
| Journal of Accounting Research | ACCOUNT | 4 | 40 | 377 | 30 | 53 | 37 | 38 | 56 | 21 | 39 | 25 | 299 | 24 |
| Accounting Review | ACCOUNT | 4 | 75 | 808 | 31 | 62 | 46 | 41 | 37 | 22 | 27 | 38 | 304 | 25 |
| International Organization* | | | 29 | 210 | 40 | 44 | 32 | 65 | 31 | 26 | 51 | 15 | 304 | 26 |
| Brookings Papers on Economic Activity* | ECON | 3 | 12 | 140 | 111 | 25 | 10 | 101 | 29 | 18 | 31 | 1 | 326 | 27 |
| Omega* | OR&MANSCI | 3 | 91 | 970 | 47 | 24 | 40 | 31 | 20 | 33 | 46 | 86 | 327 | 28 |
| Information Systems Research | INFO MAN | 4 | 63 | 960 | 13 | 59 | 28 | 51 | 61 | 35 | 49 | 37 | 333 | 29 |
| Marketing Science | MKT | 4 | 62 | 688 | 38 | 63 | 65 | 54 | 55 | 14 | 18 | 30 | 337 | 30 |
| Organizational Behavior and Human Decision Processes | PSYCH | 4 | 74 | 569 | 29 | 30 | 33 | 55 | 94 | 41 | 28 | 27 | 337 | 31 |
| Entrepreneurship Theory and Practice | ENT-SMBUS | 4 | 113 | 759 | 72 | 46 | 34 | 26 | 38 | 38 | 56 | 55 | 365 | 32 |
| Long Range Planning* | STRAT | 3 | 40 | 716 | 93 | 68 | 27 | 2 | 6 | 23 | 112 | 61 | 392 | 33 |
| Academy of Management Perspectives | | | 24 | 344 | 50 | 35 | 38 | 44 | 59 | 55 | 81 | 40 | 402 | 34 |
| Journal of Financial and Quantitative Analysis | FINANCE | 4 | 48 | 479 | 57 | 89 | 75 | 88 | 47 | 25 | 19 | 20 | 420 | 35 |
| International Journal of Management Reviews* | GEN MAN | 3 | 33 | 474 | 119 | 41 | 26 | 11 | 10 | 61 | 108 | 45 | 421 | 36 |
| Journal of Cleaner Production* | | | 709 | 3617 | 52 | 16 | 30 | 28 | 40 | 103 | 14 | 138 | 421 | 37 |
| Journal of Public Administration Research and Theory* | PUB SEC | 4 | 37 | 425 | 89 | 33 | 43 | 67 | 49 | 37 | 62 | 47 | 427 | 38 |
| Journal of Labor Economics* | ECON | 3 | 32 | 172 | 58 | 78 | 63 | 115 | 63 | 19 | 37 | 8 | 441 | 39 |
| Accounting, Organizations and Society | ACCOUNT | 4 | 38 | 465 | 48 | 69 | 35 | 34 | 42 | 72 | 79 | 66 | 445 | 40 |

**Table 4 Top 40 Journals Ranked by Sum of the Ranks (those with an * are not included in the FT Top-45 list)**

- Even in this relatively consistent set of journals, i.e., the top ones, there are some very significant differences in rank. For example the h-index would rank some of them as over 50 places lower, generally those with relatively few citations and documents such as *Brookings Papers* and the *Int. J. of Management Reviews*. The Eigenfactor and IPP have similar effects.
- In contrast, some journals would be ranked much higher on particular indicators, e.g., *Long Range Planning* would be 30 places higher on SNIP and IPP and *Brookings Papers* would be 1st on article influence score. This is a particularly divergent journal being ranked 111th on h-index.
- Apart from the ABS list, another influential one is the FT Top-45 list[12], particularly important in the MBA market. These are often seen as the very elite journals. We have indicated which journals are *not* included in that list but arguably should be. Some in the FT list that arguably should not be are: *Review of Accounting Studies* (120th), *Journal of Business Ethics* (114th) and *Human Resource Management* (112th).

We will now consider particular indicators. Table 5 shows a selection of journals that differ considerably (>=40 places) in their rank compared with the sum of ranks ranking. They are not necessarily from the top 40. We have been selective in choosing those which appear under several indicators. The numbers in brackets below are the total number of documents.

*h-index*

It is clear that the h-index is strongly affected by the number of documents published. The mean number of documents for higher ranked journals is 267 per year and for the lower journals is only 27 per year. The *Academy of Management Annals* is an interesting example because it is favoured by all the other indicators except the Eigenfactor. This journal is only published once per year and had only 6 papers in 2013, but each one was long and detailed and became highly cited. Other journals in a similar position with respect to the indicators are: *Management and Organization Review* (22), *Human Resource Management Journal* and *J. of Consumer Culture*. The *European J. of OR* (165) does well as it publishes many papers and is also well cited.

*IF, 5-Year IF, IPP*

These indicators are all similar in normalizing for the number of papers but nothing else. They therefore favour those journals with high citations per paper. Journals favoured by these metrics are, for example, *J. of Supply Chain Management* (31), *Management and Organization Review* (22) and *Human Resource Management Journal* (23), the latter two in contrast to the h-index. Journals doing poorly are: *J. Financial and Quantitative Analysis* (48), *European J. of OR* (165), and *J. of Conflict Resolution* (44) which have relatively more papers. Of the three, the IF and 5-IF have more in common than the IPP.

[12] http://www.ft.com/cms/s/2/3405a512-5cbb-11e1-8f1f-00144feabdc0.html#axzz3t47qgC3G

*SNIP*

This indicator normalizes for the citation density in the field as well as the papers published by correcting the IPP for the length of reference lists. It is difficult to see from this data to what extent it works – it would need a large sample of journals from more diverse research fields, especially science and humanities. We can see that four of the favoured journals are the same as the IPP. In fact, there is one that is significantly different – the *J. Supply Chain Management*. Interestingly, in the lower journals there are two from the supply chain area and two from the business ethics area suggesting that possibly those fields have larger reference lists which has led to their IPP being reduced.

*SJR, Eigenfactor and article influence score*

These indicators all include prestige although the Eigenfactor does not normalize for number of papers. We would therefore expect that SJR and AIS were quite similar and indeed they are. In Table 5 we can see several journals that are ranked highly by both, for example *Academy of Mgt. Annals, Quantitative Marketing and Economics, Strategic Entrepreneurship J.* and *J. of Industrial Economics*. Similarly with those downgraded - *J. Business Research, J. Business Ethics* and *Int. Marketing Review*. It is not possible to check, easily, whether these actually do differ in the prestige of the citing journals. The Eigenfactor, because of its lack of normalisation, grades several of these journals in the opposite direction. As noted above, metrics such as SJR and SNIP are difficult to investigate because of their lack of transparency and reproducibility.

Table 4 also reveals some interesting contrasts between the indicators. For example, *MIS Quarterly* is ranked very highly by the impact factors and by SNIP, but much lower by SJR and AIS. This would imply that it gets a lot of citations but from relatively less prestigious journals. However, this may be because most of its citations would come from IS journals which would themselves be less prestigious than general management journals. In contrast, the *J. of Financial Economics* and the *J. of Financial Studies* are ranked higher by the prestige indicators than the pure citation ones. This could reflect the fact that several finance journals are all ranked very highly (three in the top ten) which could mean that finances journals as a whole are highly cited and so have greater prestige as a field.

| | H-Index | IF | 5-IF | IPP | SNIP | SJR | Eigen | AIS |
|---|---|---|---|---|---|---|---|---|
| Journals ranked higher than the sum of ranks by more than 40 places | Mgt. Science<br>European J. of OR<br>J Business Research<br>California Mgt. Review<br>J. of Retailing<br>Int J of Proj. Mgt<br>J. Business Ethics | Academy of Mgt. Annals<br>Family Business Review<br>J. Supply Chain Mgt.<br>Mgt. and Organization Review<br>Strategic Entrepreneurship J<br>Human Res. Mgt. J<br>J. of Industrial Economics<br>Int. Mkting. Review<br>J. Economic Inequality<br>Int J of Proj. Mgt | Academy of Mgt. Annals<br>Research in Org. Beh.<br>J. Supply Chain Mgt.<br>Mgt. and Organization Review<br>Strategic Entrepreneurshi p J<br>J. of Industrial Economics<br>. | Academy of Mgt. Annals<br>J. Supply Chain Mgt.<br>Business Horizons<br>Human Res. Mgt. J<br>Int J of Proj. Mgt | Academy of Mgt. Annals<br>Business Horizons<br>ACM Trans Inf. Syst.<br>Int. J. of Proj. Mgt.<br>J of Economic Inequality<br>J. Professional Issues in Engineering<br>Human Res. Mgt. J<br>J. of Consumer Culture | Academy of Mgt. Annals<br>Quantitative Mkting. and Economics<br>Strategic Entrepreneurshi p J.<br>Human Res. Mgt. J.<br>J. of Industrial Economics<br>J. Risk and Uncertainty | Quantitative Mkting. and Economics<br>European J. of OR<br>Strategic Entrepreneurs hip J<br>J. of Industrial Economics<br>J. Business Research<br>J. Business Ethics | Academy of Mgt. Annals<br>Quantitative Mkting. and Economics<br>Strategic Entrepreneurship J<br>J. of Industrial Economics<br>J. Economic Inequality<br>J. of Consumer Culture<br>Int J of Proj. Mgt<br>Mathematical Finance<br>J. Risk and Uncertainty |
| Journals ranked lowerthan the sum of ranks by more than 40 places | Academy of Mgt. Annals<br>Business \|Horizons<br>Quantitative Mkting. and Economics<br>Strategic Entrepreneurship J<br>J. of Consumer Culture<br>Mgt. and Organization Review<br>Human Res. Mgt. J<br>J. of Industrial Economics<br>J. Economic Inequality | J. Financial and Quantitative Analysis<br>European J. of OR<br>Information and Mgt.<br>Int. J. Research in Mkting.<br>Business Horizons<br>J. Business Research<br>J. of Conflict Resolution<br>Mathematical Finance | J. Financial and Quantitative Analysis<br>European J. of OR<br>IMF Economic Review<br>J. of Conflict Resolution<br>Int J of Proj. Mg<br>Human Res. Mgt. J<br>Int. Mkting. Review<br>Mathematical Finance | J. of Conflict Resolution<br>J. Financial and Quantitative Analysis<br>Mathematical Finance<br>J. Risk and Uncertainty | J. Supply Chain Mgt.<br>Supply Chain Mgt.<br>Business Ethics Quarterly<br>J. of Conflict Resolution<br>Quantitative Mkting. and Economics<br>J. Business Ethics<br>J. Risk and Uncertainty | J. Business Research<br>J. Business Ethics<br>Int. Mkting. Review<br>J. Economic Inequality | Int. Mkting. Review<br>J. Supply Chain Mgt<br>Mgt. and Organization Review<br>J. of Consumer Culture | Human Res. Mgt. J<br>J. Business Research<br>J. Business Ethics<br>Int. Mkting. Review<br>European J. of OR<br>Int J of Proj. Mgt |

**Table 5 Journals that Differ in Rank Considerably Across Indicators**

### *4.3. Summary*

Reviewing the above results, we can come to the following tentative conclusions, based obviously on this particular sample of data.

- At first sight, through the correlation analysis, the indicators all appear to be very similar with very high correlation coefficient values. However, looking in more detail we see that in fact they differ considerably and individual journals may well change their rank position by over a hundred places from one indicator to another. This is very significant especially given the increasing concern with the quality of the journal as an (illicit) measure of the quality of a paper within it. It could easily lead to a journal (and also one of its papers) being classified as either a top journal or merely a low one (4* or a 2* in ABS terms).
- We can see that the theoretical differences also reveal themselves in the empirical data.
  - The total citation metrics – h-index, total cites and Eigenfactor – favour journals that publish many papers and consequently disfavour journals publishing a few, highly-cited papers. They do not normalize for field.
  - The mean citation metrics – IF, 5-IF, IPP, SNIP – favour journals that publish relatively few, highly-cited, papers and disfavour journals publishing a lot of papers, even if highly cited. Apart from SNIP, they do not normalize for field.
  - There is only limited evidence, on this sample, that SNIP normalizes significantly for field.
  - There is some evidence that the prestige indicators – SJR, AIS – do have an effect but this may be field-related rather than journal-related.
- We do not see any one indicator as superior to the others – they all have their weaknesses and biases. However, given that they are being used and that one had to make a recommendation at this point in time, we would suggest using both the h-index, because it is transparent, easy to understand and robust to poor data especially if Google Scholar is being used; and SNIP as it aims to normalizes for the number of publications and potentially the field as well[13]. There are ways of normalising the h-index for field (Iglesias & Pecharromán, 2007) although this is an area for further research (Glänzel et al., 2011).

## 5. Comparing Journal Indicators with Peer Review Journal Lists

In practice, at the moment, most journal ranking is actually done through peer reviewed lists such as the ABS list, or the Australian Business Dean's Council (ABDC) list[14] which itself is a development of the more extensive Excellence in Research for Australia (ERA) list[15] (Hall, 2011), although these may include some use of bibliometric indicators in their compilation. Interestingly, the ERA ranked list was discontinued after 2010 and now all that is available is an unranked list of the journals that were submitted in the ERA. We have already demonstrated the important effects that these lists can have on universities, departments and even individual scholars despite the extensive criticism of such lists (Adler & Harzing, 2009; Moosa, 2011; Nkomo, 2009). There have been several studies within information systems of the use of metrics (Straub, 2006; Straub & Anderson, 2010) and comparisons with expert rankings (Lowry et al., 2013; Lowry et al., 2004). Note also the San Francisco Declaration on Research Assessment (DORA, http://www.ascb.org/dora/) and the Leiden Manifesto for Research Metrics (http://www.leidenmanifesto.org/) which both set out guidelines for the proper use of metrics in evaluating research.

[13] Although the evidence for its normalizing powers is not that great in this small sample. Further research is needed here.

[14] http://www.abdc.edu.au/pages/abdc-journal-quality-list-2013.html

In the UK, the ABS list is predominant despite intense criticism (Hoepner & Unerman, 2009; Hussain, 2011; Hussain, 2013; Mingers & Willmott, 2013; Morris et al., 2009; Willmott, 2011). The main criticisms of the ABS list are: first, that the 4* journals are dominated by traditional, US-operated, largely positivistic, journals at the expense of more eclectic and innovative European and non–US ones. Second, that the distribution of 4* journals across fields is highly unequal – 42% of psychology journals but less than 5% in fields such as operations management, operational research and information systems/management (IS/IM), and none in ethics/government and management education. Third, that the coverage of journals across fields was dominated by reference disciplines (psychology, economics and social science accounted for 30% of the list). For these reasons, it is valuable to compare rankings based on indicators with the ABS rankings.

[15] http://lamp.infosys.deakin.edu.au/era/?page=jmain

| Standard Title | Field | ABS 2009 1* to 4* | Sum ranked |
|---|---|---|---|
| Brookings Papers on Economic Activity | ECON | 3 | 27 |
| Omega | OR&MANSCI | 3 | 28 |
| Long Range Planning | STRAT | 3 | 33 |
| International Journal of Management Reviews | GEN MAN | 3 | 36 |
| Journal of Labor Economics | ECON | 3 | 39 |
| Journal of Human Resources | ECON | 3 | 41 |
| Tourism Management | TOUR-HOSP | 3 | 43 |
| European Journal of Operational Research | OR&MANSCI | 3 | 44 |
| Journal of Service Research | SECTOR | 3 | 45 |
| Transportation Research Part E: Logistics and Transportation Review | SECTOR | 3 | 47 |
| Technovation | INNOV | 2 | 49 |
| Journal of Information Technology | INFO MAN | 3 | 50 |
| Human Resource Management Review | HRM&EMP | 2 | 54 |
| Family Business Review | ENT-SMBUS | 2 | 55 |
| Decision Support Systems | INFO MAN | 3 | 56 |
| Information and Management | INFO MAN | 3 | 57 |
| Journal of World Business | IB&AREA | 3 | 60 |
| International Journal of Research in Marketing | MKT | 3 | 61 |
| Journal of Management Information Systems | INFO MAN | 3 | 63 |
| Business Ethics Quarterly | ETH-GOV | 3 | 66 |
| Journal of Strategic Information Systems | INFO MAN | 3 | 67 |
| Journal of Supply Chain Management | OPS & TECH | 1 | 68 |
| Journal of Policy Analysis and Management | PUB SEC | 3 | 69 |
| Supply Chain Management | OPS & TECH | 3 | 70 |

**Table 6 Journals with High Indicator Values but Low ABS Rank**

Table 6 shows journals that score highly in terms of indicators, "Sum Ranked", but are not evaluated as 4* within ABS. The "Sum Ranked" column shows their position in the ranking of the sum of ranks. In the main these are ABS 3*, but *Technovation* is a 2* and the *J. of Supply Chain Management* only considered a 1*. In terms of the fields represented, we can see two from OR, five from IS/IM, and two from operations management which backs up the criticisms mentioned above.

| Standard Title | Field | ABS 2009 1* to 4* | SNIP Rank | Sum ranked |
|---|---|---|---|---|
| Business History | BUS HIST | 4 | 433 | 367 |
| IEEE Transactions on Engineering Management | INFO MAN | 4 | 267 | 179 |
| Industrial Relations | HRM&EMP | 4 | 251 | 159 |
| British Journal of Industrial Relations | HRM&EMP | 4 | 138 | 148 |
| Public Administration Review | PUB SEC | 4 | 241 | 138 |
| International Journal of Industrial Organization | ECON | 4 | 165 | 135 |
| Work, Employment and Society | HRM&EMP | 4 | 123 | 124 |
| Review of Accounting Studies | ACCOUNT | 4 | 177 | 120 |
| Human Resource Management | HRM&EMP | 4 | 196 | 112 |
| Journal of Risk and Uncertainty | SOC SCI | 4 | 231 | 107 |
| Journal of Retailing | MKT | 4 | 130 | 96 |
| British Journal of Management | GEN MAN | 4 | 116 | 79 |
| Journal of Product Innovation Management | INNOV | 4 | 67 | 64 |
| Journal of Vocational Behavior | HRM&EMP | 4 | 117 | 59 |
| Annals of Tourism Research | TOUR-HOSP | 4 | 70 | 58 |
| Human Relations | ORG STUD | 4 | 90 | 52 |
| Leadership Quarterly | ORG STUD | 4 | 110 | 51 |
| Journal of Occupational and Organizational Psychology | PSYCH | 4 | 95 | 48 |
| Organization Studies | ORG STUD | 4 | 105 | 42 |
| Accounting, Organizations and Society | ACCOUNT | 4 | 42 | 40 |
| Journal of Public Administration Research and Theory | PUB SEC | 4 | 49 | 38 |
| Journal of Financial and Quantitative Analysis | FINANCE | 4 | 47 | 35 |
| Entrepreneurship Theory and Practice | ENT-SMBUS | 4 | 38 | 32 |
| Organizational Behavior and Human Decision Processes | PSYCH | 4 | 94 | 31 |

**Table 7 Journals with Low Indicator Values but High ABS Rank**

Table 7 shows the opposite, namely journals ranked as 4* in ABS but being relatively lowly ranked in terms of indicators. This Table does not show a preponderance of US journals or dominance of fields like Psychology. In fact, most of the journals are non-US and several come from a particular field, HR (5) and organization studies (3). To check whether this was a problem with low density fields the Table also show the SNIP rank (which aims to correct for this) but in the main the SNIP ranks are even worse. *Business History* is a particular outlier being ranked only 367$^{th}$.

| Field | Mean ABS score | Rank of ABS score | Mean Sum Ranked |
|---|---|---|---|
| PSYCH Average | 2.71 | 2 | 110.14 |
| OR&MANSCI Average | 2.75 | 1 | 190.50 |
| PUB SEC Average | 2.60 | 4 | 202.10 |
| INFO MAN Average | 2.35 | 10 | 232.60 |
| HRM&EMP Average | 2.64 | 3 | 239.79 |
| GEN MAN Average | 2.41 | 7 | 259.33 |
| OPS MGT & TECH Average | 1.88 | 19 | 262.31 |
| ORG STUD Average | 2.38 | 9 | 269.38 |
| SOC SCI Average | 2.29 | 11 | 272.53 |
| ECON Average | 2.28 | 12 | 272.80 |
| INNOV Average | 2.00 | 16 | 286.25 |
| FINANCE Average | 2.50 | 5 | 293.73 |
| ENT-SMBUS Average | 2.21 | 13 | 294.13 |
| TOUR-HOSP Average | 1.90 | 18 | 299.80 |
| ETH-GOV Average | 1.80 | 20 | 301.90 |
| STRAT Average | 2.20 | 14 | 310.60 |
| SECTOR Average | 1.55 | 23 | 350.36 |
| MKT Average | 1.95 | 17 | 353.88 |
| ACCOUNT Average | 2.41 | 8 | 391.18 |
| LAW Average | 2.00 | 15 | 404.00 |
| IB&AREA Average | 1.72 | 21 | 405.22 |
| MGDEV&ED Average | 1.71 | 22 | 440.93 |
| BUS HIST Average | 2.43 | 6 | 523.43 |

**Table 8 Comparing Fields in ABS and Indicator Ranking**

In Table 8 we look at the fields. The second and fourth columns show the mean scores per field for the ABS grade and the sum ranked respectively. The Table is sorted in terms of sum ranked but the third column shows the ranks of the ABS mean. The rank correlation is 0.61. We can see that fields like information management and operations management do poorly in ABS while business history, accounting and finance do relatively well. This result agrees with other research concerning these fields. Templeton and Lewis (2015) compared the prestige of B&M journals within AACSB business schools in terms of how highly they were valued, based on surveys of the Schools, compared with their citation performance based on a range of metrics (similar to ours). They found that information systems especially, but also operations management and quantitative methods were undervalued in comparison with the citation performance of the subjects' journals. Valacich et al (2006) found that the publications opportunities in top journals was limited for IS researchers, i.e., there were relatively few IS journals considered to be elite, and they published relatively few papers. Lowry et al (2013) compared expert opinion (from a survey) and bibliometric measures for IS journals only and found that the results were extremely similar so that suggested that metrics should be used instead in the future.

The overall conclusions of this section is that there are systematic discrepancies between the ABS list and rankings based on citation indicators which leads to questions about what justification there is for these differences.

## 6. Conclusions

This paper has considered the main journal impact indicators that are currently available through citation databases as these are the primary ones that are used in practice for decisions about journal ranking lists, destinations for research papers, jobs, promotions, and submissions to research evaluation programmes.

There are several general issues to be noted in terms of the appropriate use of these metrics. First, citation data is always highly skewed and this calls into question the validity of measures based on parameters such as the mean, especially for the evaluation of individual cases of journals or researchers. Second, there is the ecological fallacy of making judgements about individuals on the bases of whole population characteristics. For example, judging the quality of individual papers purely in terms of the journal they are published in, or judging individuals in terms of particular journals when they publish across a range, potentially in different fields. Third, there is the whole issue of using the number of citations, especially in the form of short term impact factors, as a measure of journal quality anyway, certainly in the social sciences.

In comparing these particular metrics, we have found that, at first sight, they appear to be highly correlated and that this may lead users to believe that they are in fact very similar in their results and that it does not, perhaps, matter too much which ones are used. But, in fact, these correlations mask significant differences between them, both theoretically and empirically, and these differences can have major effects on the rankings of individual journals. Given the extensive use of journal ranking lists and journal metrics in research evaluation, and the consequences this can have on departments and individuals, it is important that these effects are recognized and factored into any decisions being made.

The differences occur because of the inevitable biases in any form of metric dependent on the particular underlying assumptions and manner of its calculation. The main theoretical differences between the indicators are: whether they normalize for the number of papers generating the citations, and the subject area or field; whether they take into consideration the prestige of the citing journals; whether they are affected by skewed data; whether they are transparent, easily interpretable and robust to poor data; whether they are reliant on a particular proprietary database; and whether they are transparent and reproducible by other researchers. These theoretical differences were largely corroborated in the empirical comparisons.

We also compared rankings formed on the basis of the citation metrics with a well-known journal list that is used extensively within research assessments. Many instances were found where journals that performed well in terms of citations were ranked relatively lowly and journals that were ranked highly had little citation impact.

In terms of practical recommendations, we do not feel that any one indicator stands out as superior at this time, they all have their limitations. Equally, however, peer review and expert journal lists are subjective and biased in many ways. We feel therefore that the best approach is to employ several metrics along with peer review if it is really felt necessary to produce ranked lists of journals but then to exercise great caution in inferring from the general high-level results down to the performance of individuals. If we were to recommend any metrics, we would suggest SNIP, which normalizes for papers and also field (although this was not very evident on our data), and the h-index which is transparent, easy to understand, and robust to poor data thus being especially useful with Google Scholar.

This particular study does have significant limitations: it was conducted only within one disciplinary field, business and management, although that is a very diverse field which displays many of the characteristics of social science as a whole. It was also limited in terms of the number of journals that could actually be included in the final analyses because of limitations in some of the data sources.

We feel that further research is needed, particular in the following areas: i) Large scale tests of different forms of normalisation, both citing- and cited-side, and a priori (i.e., adjusting the citations before an indicator is calculated) and a posteriori (adjusting the indicator after it is calculated) (Glänzel et al., 2011). ii) Investigating ways of normalizing Google Scholar data and improving its quality. iii) Investigating the possibilities of creating weighted aggregated indices that might overcome the limitations of any particular one (Ennas et al., 2015). iv) Investigating indicators that are not currently supported by WoS or Scopus such as I3 which avoids the problem of skewness.

## References

Aad, G., Abajyan, T., Abbott, B., Abdallah, J., Abdel Khalek, S., Abdelalim, A. A., . . . Zwalinski, L. (2012). Observation of a new particle in the search for the Standard Model Higgs boson with the ATLAS detector at the LHC. *Physics Letters B, 716*(1), 1-29.

Abramo, G., & D'Angelo, C. (2011). Evaluating research: from informed peer review to bibliometric. *Scientometrics, 87*(3), 499-514.

Adler, N., & Harzing, A.-W. (2009). When knowledge wins: Transcending the sense and nonsense of academic rankings. *Academy of Management Learning and Education, 8*(1), 72-95.

Adriaanse, L., & Rensleigh, C. (2013). Web of Science, Scopus and Google Scholar. *The Electronic Library, 31*(6), 727-744.

Alonso, S., Cabrerizo, F. J., Herrera-Viedma, E., & Herrera, F. (2009). h-Index: A review focused in its variants, computation and standardization for different scientific fields. *Journal of Informetrics, 3*(4), 273-289.

Amara, N., & Landry, R. (2012). Counting citations in the field of business and management: why use Google Scholar rather than the Web of Science. *Scientometrics, 93*(3), 553-581.

Association of Business Schools. (2010, 1/3/2010). Academic journal quality guide, from http://www.bizschooljournals.com/node/4

Bergstrom, C. (2007). Measuring the value and prestige of scholarly journals. *Coll Res Libr News, 68*(5), 3146.

Bollen, J., Rodriquez, M. A., & Van de Sompel, H. (2006). Journal status. *Scientometrics, 69*(3), 669-687.

Bornman, L., & Daniel, H.-D. (2005). Does the h-index for ranking of scientists really work? *Scientometrics, 65*(3), 391-392.

Bornmann, L., & Daniel, H.-D. (2007). What do we know about the h index? *Journal of the American Society for Information Science and Technology, 58*(9), 1381-1385.

Bornmann, L., Leydesdorff, L., & Mutz, R. (2013). The use of percentiles and percentile rank classes in the analysis of bibliometric data: Opportunities and limits. *Journal of Informetrics, 7*(1), 158-165.

Bornmann, L., & Marx, W. (2015). Methods for the generation of normalized citation impact scores in bibliometrics: Which method best reflects the judgement of experts? *Journal of Informetrics, 9*(2), 408-418.

Bornmann, L., Mutz, R., & Daniel, H. D. (2008). Are there better indices for evaluation purposes than the h index? A comparison of nine different variants of the h index using data from biomedicine. *Journal of the American Society for Information Science and Technology, 59*(5), 830-837.

Bornmann, L., Thor, A., Marx, W., & Schier, H. (2016). The application of bibliometrics to research evaluation in the humanities and social sciences: An exploratory study using normalized Google Scholar data for the publications of a research institute. *Journal of the Association for Information Science and Technology*, n/a-n/a.

Brumback, R. (2008). Worshipping false idols: the impact factor dilemma. *Journal Child Neurology, 23*, 365-367.

Campanario, J. M. (2011). Empirical study of journal impact factors obtained using the classical two-year citation window versus a five-year citation window. *Scientometrics, 87*(1), 189-204.

Chartered Association of Business Schools. (2015). Academic journal guide 2015, from http://charteredabs.org/academic-journal-guide-2015/

Costas, R., & Bordons, M. (2007). The h-index: Advantages, limitations and its relation with other bibliometric indicators at the micro level. *Journal of Informetrics, 1*, 193-203.

Cronin, B. (2001). Hyperauthorship: A postmodern perversion or evidence of a structural shift in scholarly communication practices? *Journal of the American Society for Information Science and Technology, 52*(7), 558-569.

Cronin, B., & Sugimoto, C. (Eds.). (2014). *Beyond Bibliometrics: Harnessing Multidimensional Indicators nof Scholarly Impact*. London: MIT Press.
Egghe, L. (2006). Theory and practice of the g-index. *Scientometrics*, *69*(1), 131-152.
Ennas, G., Biggio, B., & Di Guardo, M. (2015). Data-driven journal meta-ranking in business and management. *Scientometrics*, 1-19.
Evidence Ltd. (2007). The use of bibliometrics to measure research quality in UK higher education institutions. Leeds: Evidence Ltd. http://www.hefce.ac.uk/pubs/rdreports/2007/rd19%5F07/.
Falagas, M. E., Kouranos, V. D., Arencibia-Jorge, R., & Karageorgopoulos, D. E. (2008). Comparison of SCImago journal rank indicator with journal impact factor. *The FASEB Journal*, *22*(8), 2623-2628.
Fersht, A. (2009). The most influential journals: Impact Factor and Eigenfactor. *Proceedings of the National Academy of Sciences*, *106*(17), 6883-6884.
Franceschet, M. (2010). A comparison of bibliometric indicators for computer science scholars and journals on Web of Science and Google Scholar. *Scientometrics*, *83*(1), 243-258.
Franceschini, F., & Maisano, D. A. (2010). Analysis of the Hirsch index's operational properties. *European Journal of Operational Research*, *203*(2), 494-504.
García-Pérez, M. A. (2010). Accuracy and completeness of publication and citation records in the Web of Science, PsycINFO, and Google Scholar: A case study for the computation of h indices in Psychology. *Journal of the American Society for Information Science and Technology*, *61*(10), 2070-2085.
Garfield, E. (1955). Citation Indexes for Science: A New Dimension in Documentation through Association of Ideas. *Science*, *122*(3159), 108-111.
Garfield, E., & Sher, I. H. (1963). New factors in the evaluation of scientific literature through citation indexing. *American Documentation*, *14*(3), 195-201.
Giles, L., & Khabsa, M. (2014). The Number of scholarly documents on the Web. *PLoS ONE*, *9*(5), e93949.
Glänzel, W. (2006). On the h-index - a mathematical approach to a new measure of publication activity and citation impact. *Scientometrics*, *67*(2), 315-321.
Glänzel, W., & Moed, H. K. (2002). Journal impact measures in bibliometric research. *Scientometrics*, *53*(2), 171-193.
Glänzel, W., Schubert, A., Thijs, B., & Debackere, K. (2011). A priori vs. a posteriori normalisation of citation indicators. The case of journal ranking. *Scientometrics*, *87*(2), 415-424.
González-Pereira, B., Guerrero-Bote, V. P., & Moya-Anegón, F. (2010). A new approach to the metric of journals' scientific prestige: The SJR indicator. *Journal of Informetrics*, *4*(3), 379-391.
Gross, P., & Gross, E. (1927). College libraries and chemical education. *Science* *66*(385-389).
Guerrero-Bote, V. P., & Moya-Anegón, F. (2012). A further step forward in measuring journals' scientific prestige: The SJR2 indicator. *Journal of Informetrics*, *6*(4), 674-688.
Hall, C. (2011). Publish and perish? Bibliometric analysis, journal ranking and the assessment of research quality in tourism. *Tourism Management*, *32*(1), 16-27.
Harzing, A.-W. (2009). Journal Quality List 55th. Retrieved January, 2015, from http://www.harzing.com/jql.htm
Harzing, A.-W., & van der Wal, R. (2008). Google Scholar as a new source for citation analysis. *Ethics in Science and Environmental Politics*, *8*, 61-73.
Harzing, A.-W., & Van der Wal, R. (2009). A Google Scholar h-index for journals: An alternative metric to measure journal impact in economics and business? *J. Am. Soc. for Information Science and Technology*, *60*(1), 41-46.
HEFCE. (2008). Counting what is measured or measuring what counts: HEFCE. http://www.hefce.ac.uk/pubs/hefce/2007/07_34/.
Hicks, D. (2012). Performance-based university research funding systems. *Research policy*, *41*(2), 251-261.
Hirsch, J. (2005). An index to quantify an individual's scientific research output. *Proceedings of the National Academy of Sciences of the United States of America*, *102*(46), 16569-16572.
Hoepner, A., & Unerman, J. (2009). Explicit and implicit subject bias in the ABS Journal Quality Guide. *Accounting Education*, *21*(1), 3-15.

Hussain, S. (2011). Food for thought on the ABS academic journal quality guide. *Accounting Education, 20*(6), 545-559.

Hussain, S. (2013). Journal Fetishism and the 'Sign of the 4' in the ABS guide: A Question of Trust? *Organization, online*.

Iglesias, J., & Pecharromán, C. (2007). Scaling the h-index for different scientific ISI fields. *Scientometrics, 73*(3), 303-320.

Jacso, P. (2005). As we may search - Comparison of major features of the Web of Science, Scopus, and Google Scholar citation-based and citation-enhanced databases. *Current Science, 89*(9), 1537-1547.

Jin, B., Liang, L., Rousseau, R., & Egghe, L. (2007). The R- and AR-indices: Complementing the h-index. *Chinese Science Bulletin, 52*(6), 855-863.

Lancho-Barrantes, B. S., Guerrero-Bote, V. P., & Moya-Anegón, F. (2010). What lies behind the averages and significance of citation indicators in different disciplines? *Journal of Information Science, 36*(3), 371-382.

Larivière, V., Archambault, É., Gingras, Y., & Vignola-Gagné, É. (2006). The place of serials in referencing practices: Comparing natural sciences and engineering with social sciences and humanities. *Journal of the American Society for Information Science and Technology, 57*(8), 997-1004.

Lehmann, S., Jackson, A. D., & Lautrup, B. E. (2006). Measures for measures. [10.1038/4441003a]. *Nature, 444*(7122), 1003-1004.

Leydesdorff, L. (2012). Alternatives to the journal impact factor: I3 and the top-10% (or top-25%?) of the most highly cited papers. *Scientometrics, 92*, 355-365.

Leydesdorff, L., & Bornmann, L. (2011). Integrated impact indicators compared with impact factors: An alternative research design with policy implications. *Journal of the American Society for Information Science and Technology, 62*(11), 2133-2146.

Leydesdorff, L., & Bornmann, L. (2014). The Operationalization of "Fields" as WoS Subject Categories (WCs) in Evaluative Bibliometrics: The cases of" Library and Information Science" and" Science & Technology Studies". *arXiv preprint arXiv:1407.7849*. Retrieved from

Leydesdorff, L., Bornmann, L., Mutz, R., & Opthof, T. (2011). Turning the tables on citation analysis one more time: Principles for comparing sets of documents. *Journal of the American Society for Information Science and Technology, 62*(7), 1370-1381.

Leydesdorff, L., de Moya-Anegón, F., & de Nooy, W. (2014). Aggregated journal-journal citation relations in Scopus and Web-of-Science matched and compared in terms of networks, maps, and interactive overlays. *arXiv preprint arXiv:1404.2505*.

Leydesdorff, L., & Milojević, S. (2015). The Citation Impact of German Sociology Journals: Some Problems with the Use of Scientometric Indicators in Journal and Research Evaluations. *Soziale Welt, 66(2), 193-204. , 66*(2), 193-204.

Leydesdorff, L., & Opthof, T. (2010). Scopus's source normalized impact per paper (SNIP) versus a journal impact factor based on fractional counting of citations. *Journal of the American Society for Information Science and Technology, 61*(11), 2365-2369.

Lowry, P. B., Moody, G. D., Gaskin, J., Galletta, D. F., Humpherys, S. L., Barlow, J. B., & Wilson, D. W. (2013). EVALUATING JOURNAL QUALITY AND THE ASSOCIATION FOR INFORMATION SYSTEMS SENIOR SCHOLARS' JOURNAL BASKET VIA BIBLIOMETRIC MEASURES: DO EXPERT JOURNAL ASSESSMENTS ADD VALUE? [Article]. *MIS quarterly, 37*(4), 993-A921.

Lowry, P. B., Romans, D., & Curtis, A. (2004). Global journal prestige and supporting disciplines: A scientometric study of information systems journals. *Journal of the Association for Information Systems, 5*(2), 29-75.

Ma, N., Guan, J., & Zhao, Y. (2008). Bringing PageRank to the citation analysis. *Information Processing and Management, 44*, 800-810.

Mahdi, S., D'Este, P., & Neely, A. (2008). Citation counts: Are they good predictors of RAE scores? London: AIM Research.

Meho, L., & Rogers, Y. (2008). Citation counting, citation ranking, and h-index of human-computer interaction researchers: A comparison of Scopus and Web of Science. *Journal American Society for Information Science and Technology, 59*(11), 1711-1726.

Meho, L., & Yang, K. (2007). Impact of data sources on citation counts and rankings of LIS faculty: Web of Science, Scopus and Google Scholar. *Journal American Society for Information Science and Technology, 58*(13), 2105-2125.

Mingers, J. (2008). Measuring the research contribution of management academics using the Hirsch-index. *Journal Operational Research Society, 60*(8), 1143-1153.

Mingers, J. (2014). Problems with SNIP. *Journal of Informetrics, 8*(4), 890-894.

Mingers, J., & Leydesdorff, L. (2015a). Identifying research fields within business and management: A journal cross-citation analysis. *Journal of the Operational Research Society, 66*(8), 1370-1384.

Mingers, J., & Leydesdorff, L. (2015b). A review of theory and practice in scientometrics. *European Journal of Operational Research, 246*(1), 1-19.

Mingers, J., & Lipitakis, E. (2010). Counting the citations: A comparison of Web of Science and Google Scholar in the field of management. *Scientometrics 85*(2), 613-625.

Mingers, J., Macri, F., & Petrovici, D. (2012). Using the h-index to measure the quality of journals in the field of Business and Management. *Information Processing & Management, 48*(2), 234-241.

Mingers, J., & Willmott, H. (2013). Taylorizing business school research: On the "one best way" performative effects of journal ranking lists. *Human Relations, 66*(8), 1051-1073.

Moed, H. (2000). Bibliometric indicators reflect publication and management strategies. *Scientometrics, 47*(2), 323-346.

Moed, H. (2007). The future of research evaluation rests with an intelligent combination of advanced metrics and transparent peer review. *Science and Public Policy, 34*(8), 575-583.

Moed, H. (2010a). Measuring contextual citation impact of scientific journals. *Journal of Informetrics, 4*(3), 265-277.

Moed, H. (2010b). The Source-Normalized Impact per Paper (SNIP) is a valid and sophisticated indicator of journal citation impact. *arXiv preprint*. Retrieved from

Moed, H. (2015). Comprehensive indicator comparisons intelligible to non-experts: The case of two SNIP versions. Retrieved from

Moed, H., De Bruin, R., & Van Leeuwen, T. (1995). New bibliometric tools for the assessment of national research performance: Database description, overview of indicators and first applications. *Scientometrics, 33*(3), 381-422.

Moed, H., & Visser, M. (2008). Appraisal of Citation Data Sources. Leiden: Centre for Science and Technology Studies, Leiden University. http://www.hefce.ac.uk/pubs/rdreports/2008/rd17_08/.

Moed, H. F. (2010c). CWTS crown indicator measures citation impact of a research group's publication oeuvre. *Journal of Informetrics, 4*(3), 436-438.

Moosa, I. (2011). The demise of the ARC journal ranking scheme: an ex post analysis of the accounting and finance journals. *Accounting & Finance, 51*(3), 809-836.

Morris, M., Harvey, C., & Kelly, A. (2009). Journal rankings and the ABS Journal Quality Guide. *Management Decision, 47*(9), 1441-1451.

Moussa, S., & Touzani, M. (2010). Ranking marketing journals using the Google Scholar-based hg-index. *Journal of Informetrics, 4*, 107-117.

Nkomo, S. M. (2009). The Seductive Power of Academic Journal Rankings: Challenges of Searching for the Otherwise. [Article]. *Academy of Management Learning & Education, 8*(1), 106-112.

Norris, M., & Oppenheim, C. (2010). The h-index: a broad review of a new bibliometric indicatornull. *Journal of Documentation, 66*(5), 681-705.

Northcott, D., & Linacre, S. (2010). Producing spaces for academic discourse:The impact of research assessment exercises and journal quality rankings. *Australian Accounting Review, 20*(1), 38-54.

Oppenheim, C. (2007). Using the h-index to rank influential British researchers in Information Science and Librarianship. *Journal of the American Society for Information Science and Technology, 58*(2), 297-301.

Page, L., Brin, S., Motwani, R., & Winograd, T. (1999). The PageRank citation ranking: Bringing order to the web.
Pinski, G., & Narin, F. (1976). Citation influence for journal aggregates of scientific publications: Theory, with application to the literature of physics. *Information Processing & Management, 12*(5), 297-312.
Pislyakov, V. (2009). Comparing two "thermometers": Impact factors of 20 leading economic journals according to Journal Citation Reports and Scopus. *Scientometrics, 79*(3), 541-550.
Rafols, I., Leydesdorff, L., O'Hare, A., Nightingale, P., & Stirling, A. (2012). How journal rankings can suppress interdisciplinary research: A comparison between innovation studies and business & management. *Research Policy, 41*(7), 1262-1282.
Rebora, G., & Turri, M. (2013). The UK and Italian research assessment exercises face to face. *Research policy, 42*(9), 1657-1666.
Saad, G. (2006). Exploring the h-index at the author and journal levels using bibliometric data of productive consumer scholars and business-related journals respectively. *Scientometrics, 69*(1), 117-120.
Seglen, P. O. (1992). The skewness of science. *Journal of the American Society for Information Science, 43*(9), 628-638.
Straub, D. (2006). The Value of Scientometric Studies: An Introduction to a Debate on IS as a Reference Discipline. [Article]. *Journal of the Association for Information Systems, 7*(5), 241-245.
Straub, D., & Anderson, C. (2010). Journal Quality and Citations: Common Metrics and Considerations about Their Use, Editorial, *MIS quarterly*, pp. iii-xii. Retrieved from http://search.ebscohost.com/login.aspx?direct=true&db=bth&AN=48476539&site=ehost-live
Templeton, G. F., & Lewis, B. R. (2015). Fairness in the institutional valuation of business journals. *Mis Quarterly, 39*(3), 523-539.
Todeschini, R. (2011). The j-index: a new bibliometric index and mulitvariate comparisons between other bibliometric indices. *Scientometrics, 87*, 621-639.
Truex III, D., Cuellar, M., & Takeda, H. (2009). Assessing scholarly influence: Using the Hirsch indices to reframe the discourse. *Journal of the Association for Information Systems, 10*(7), 560-594.
Valacich, J. S., Fuller, M. A., Schneider, C., & Dennis, A. R. (2006). Publication Opportunities in Premier Business Outlets: How Level Is the Playing Field? [Article]. *Information Systems Research, 17*(2), 107-125.
van Leeuwen, T., & Moed, H. (2002). Development and application of journal impact measures in the Dutch science system. *Scientometrics, 53*(2), 249-266.
van Raan, A. (2003). The use of bibliometric analysis in research performance assessment and monitoring of interdisciplinary scientific developments. *Technology Assessment - Theory and Practice, 1*(12), 20-29.
van Raan, A. (2005a). Comparison of the hirsch-index with standard bibliometric indicators and with peer judgement for 147 chemistry research groups. *Scientometrics, 67*(3), 491-502.
van Raan, A. (2005b). Fatal attraction: Conceptual and methodological problems in the ranking of universities by bibliometric methods. *Scientometrics, 62*(1), 133-143.
van Raan, A. (2005c). Measuring science: Capita selectaof current main issues. In H. Moed, W. Glenzel & U. Schmoch (Eds.), *Handbook of Quantitative Science and Technology Research* (Vol. 19-50). New York: Springer.
Walker, J., Salter, A., & Salandra, R. (2015). Initial Findings from the Survey of UK Business Academics: Henley Business School. http://hly.ac/Business-Academics
Waltman, L. (2015). A review of the literature on citation impact indicators. *Journal of Informetrics, 10*(2), 365-391.
Waltman, L., & van Eck, N. (2013). A systematic empirical comparison of different approaches for normalizing citation impact indicators. *Journal of Informetrics, 7*(4), 833-849.
Waltman, L., van Eck, N., van Leeuwen, T., & Visser, M. (2013). Some modifications to the SNIP journal impact indicator. *Journal of Informetrics, 7*(2), 272-285.

West, J., Bergstrom, T., & Bergstrom, C. (2010). Big Macs and Eigenfactor scores: Don't let correlation coefficients fool you. *Journal of the American Society for Information Science and Technology, 61*(9), 1800-1807.

Wilhite, A., & Fong, E. (2012). Coercive citation in academic publishing. *Science, 335*(6068), 542-543.

Willmott, H. (2011). Journal list fetishism and the perversion of scholarship: reactivity and the ABS list. *Organization, 18*(4), 429-442.

Wouters, P., Thelwall, M., Kousha, K., Waltman, L., de Rijcke, S., Rushforth, A., . . . Wouters, P. (2015). The Metric Tide: Report of the Independent Review of the Role of metrics in Research Assessment and Management. London: HEFCE.

Zitt, M. (2011). Behind citing-side normalization of citations: some properties of the journal impact factor. *Scientometrics, 89*(1), 329-344.

Zitt, M., & Small, H. (2008). Modifying the journal impact factor by fractional citation weighting: The audience factor. *Journal of the American Society for Information Science and Technology, 59*(11), 1856-1860.